\documentclass{iau} 
\usepackage{graphicx}
\usepackage{natbib}

\newcommand\apj{{ApJ}}
\newcommand\apjl{{ApJL}}
\newcommand\apjs{{ApJS}}
\newcommand\prc{{PRC}}
\newcommand\prd{{PRD}}
\newcommand\physrep{{Phys.~Rep.}}

\title[Neutrino Interactions from the Virial EOS] 
{Core-Collapse Supernova Simulations including Neutrino Interactions from the Virial EOS}

\author[O'Connor \emph{et al.}]   
{Evan O'Connor$^{1,2}$, C.~J. Horowitz$^3$, Zidu Lin$^3$, \and Sean Couch$^{4,5,6,7}$}

\affiliation{
$^1$Department of Astronomy and Oskar Klein Centre, Stockholm University, AlbaNova, SE-106 91 Stockholm, Sweden email: {\tt evan.oconnor@astro.su.se} \\[\affilskip]
$^2$NCSU, Department of Physics, Campus Code 8202, Raleigh, NC, 27695, USA \\[\affilskip]
$^3$Center for Exploration of Energy and Matter and Department of Physics, Indiana University, Bloomington, IN 47408, USA\\[\affilskip]
$^4$Department of Physics and Astronomy, MSU, East Lansing, MI 48824, USA\\[\affilskip]
$^5$Department of Computational Mathematics, Science, and Engineering, Michigan State University, East Lansing, MI 48824, USA\\[\affilskip]
$^6$National Superconducting Cyclotron Laboratory, MSU, East Lansing, MI 48824, USA\\[\affilskip]
  $^7$Joint Institute for Nuclear Astrophysics, MSU, East Lansing, MI 48824, USA\\[\affilskip]}

\pubyear{2017}
\volume{331}  
\jname{SN\,1987A, 30 years later}
\editors{M. Renaud, A. Marcowith, G. Dubner, A. Ray, \& A. Bykov, eds.}

\begin{document}

\maketitle

\begin{abstract}

  Core-collapse supernova explosions are driven by a central engine
  that converts a small fraction of the gravitational binding energy
  released during core collapse to outgoing kinetic energy. The
  suspected mode for this energy conversion is the neutrino
  mechanism, where a fraction of the neutrinos emitted from the newly
  formed protoneutron star are absorbed by and heat the matter
  behind the supernova shock.  Accurate neutrino-matter interaction
  terms are crucial for simulating these explosions.  In this
  proceedings for IAUS 331, \emph{SN\,1987A, 30 years later}, we
  explore several corrections to the neutrino-nucleon scattering
  opacity and demonstrate the effect on the dynamics of the
  core-collapse supernova central engine via two dimensional
  neutrino-radiation-hydrodynamics simulations. Our results reveal
  that the explosion properties are sensitive to corrections to the
  neutral-current scattering cross section at the 10-20\% level, but
  only for densities at or above $\sim 10^{12}$\,g\,cm$^{-3}$.

\end{abstract}

\firstsection 
\section{Introduction}
 
The observation of neutrinos from SN\,1987A confirmed the basic nature
of core collapse and highlighted the important role neutrinos play.
Since then, 30 years ago, the community has made tremendous progress
in elucidating this process. While neutrinos are an extreme \emph{sink} of
energy, and part of the reason that the supernova shock initially
stalls, they are also thought to be crucial in reviving the supernova
shock via the \emph{neutrino mechanism}. The neutrino mechanism relies primarily on
charged-current heating by electron neutrinos and antineutrinos on the
neutrons and protons located behind the shock. This heating gives rise
to the so-called gain region, where the heating by the background
neutrino field being emitted from deeper regions is larger than the
neutrino cooling from the matter. The net result is an increasing
internal energy and, if sufficient, the development of an explosion.
The amount of neutrino heating depends sensitively on the spectra of
the background neutrino radiation fields, both the overall
normalization (the luminosity) and its shape (mean energy, root mean
squared energy, etc.).

One of the major difficulties in modeling core-collapse supernovae is
the treatment of the neutrino radiation fields. Soon after core
bounce, in the core of the protoneutron star, the neutrinos are
completely trapped. The mean free path is much shorter than the
typical length scales and these neutrinos are in equilibrium with the
surrounding matter. However, as one moves away from the core, the
densities drop rapidly, the mean free path increases, and the
neutrinos decouple from the matter. By $\sim$100\,km, most neutrinos
are essentially free streaming and only a small percentage
($\sim5-10\%$) will have interactions before exiting the star. It is
critical to model the transition away from equilibrium since it is
this transition which sets the spectrum of the radiation field
responsible for the neutrino heating further out. To model this
transition precisely one needs spectral neutrino transport and
accurate neutrino opacities.

In this proceedings for IAU Symposium 331, ``SN 1987A, 30 years
later'', we present a series of 2D core-collapse supernova simulations
with spectral neutrino transport where we investigate in detail the
effect of varying the neutrino opacities on the core-collapse
supernova explosion mechanism. In particular, we use {\tt FLASH}
\citep{fryxell:00, couch:14a, oconnor:15b} which is outfitted with an
energy-dependent M1 neutrino transport scheme (see below) and an effective general
relativistic treatment of gravity. We explore recent corrections to
the neutral-current scattering cross sections proposed by
\cite{horowitz:17}. We do this generally for four progenitor models,
and in detail for one model in particular. Our main conclusion is that
the core-collapse supernova explosion mechanism is sensitive to the
details of the neutral current cross section at densities of
$\sim 10^{12}-10^{13}$\,g\,cm$^{-3}$. This is currently the limit
of model-independent calculations of the neutrino interactions, therefore
care and caution must be used when developing models for higher densities.

\section{Methods}
\label{sec:methods}

Our {\tt FLASH} simulations follow the methods of O'Connor \& Couch
(2017; \emph{in prep}), a significantly updated version of
\cite{oconnor:15b}. Our velocity-dependent neutrino transport scheme
is based on the M1 formalism, where we evolve the first two angular
moments of the neutrino distribution function, the energy density and
the momentum density. We evolve these moments as a function of
neutrino species (3 species in total, $\nu_e$, $\bar{\nu}_e$, and
$\nu_x=\{\nu_\mu, \bar{\nu}_\mu, \nu_\tau, \bar{\nu}_\tau\}$) and
neutrino energy (our simulations use 12 energy groups for each
species). We fully take into account gravitational red shifting. We
use the LS220 EOS \citep{lseos:91}. Our neutrino opacities come from {\tt
  NuLib}, an open-source neutrino interaction library
\citep{oconnor:15a}. The base rates are based on \cite{Bruenn:85} and
include corrections for weak magnetism and nuclear recoil
\citep{horowitz:02}. The specific corrections we explore here are based
on \cite{horowitz:17}. In \cite{horowitz:17}, we use the Virial
equation of state (EOS) to derive model-independent expressions for
the axial response function ($S_A$) of the neutral-current,
neutrino-nucleon scattering cross section, which is given as
\begin{equation}
  \frac{1}{V}\frac{d\sigma}{d\Omega} = \frac{G_F^2E^2_\nu}{16
  \pi^2}\left[g_a^2(3-\cos(\theta))(n_n+n_p)S_A +
  (1+\cos(\theta))n_nS_V\right]\,.\label{eq:sigma}
\end{equation}
The overall effect of the corrections, due to many-body effects, is to
lower the axial response function, $S_A$, and therefore lower the
total neutral-current scattering cross section.  Since the Virial EOS
is only valid at low densities, we transition to a model-dependent
expression for $S_A$ at high densities, which is based on the random
phase approximation (RPA) work of \cite{burrows:98}.  We empirically
fit our results with the following expression for use in our
simulations.
\begin{equation}
S_A^f(n,T,Y_e) = \frac{1}{1 + A (1+B e^{-C})}\,,\label{eq:S_A}
\end{equation}
where
\noindent
$$ A(n,T,Y_e)  =  A_0\frac{n(1-Y_e + Y_e^2)}{T^{1.22}}\,;\ B(T) =  \frac{B_0}{T^{0.75}}\,; \ C(n,T,Y_e)  =  C_0\frac{nY_e(1-Y_e)}{T^{0.5}} + D_0 \frac{n^4}{T^6}\,,$$
where $n$ is the baryon density in units of nucleons/fm$^3$, $T$ is
the matter temperature in MeV, and $Y_e$ is the electron fraction. We
take as constants, $A_0=920$, $B_0=3.05$, $C_0=6140$, and
$D_0=1.5\times10^{13}$. We note that setting $B_0=0$ removes the
Virial EOS contributions to the neutrino response and effectively reduces our
expressions for the axial response function to the values from
\cite{burrows:98}.  As noted in \cite{horowitz:17}, at low densities,
where the Virial EOS is valid, the Virial contributions decrease $S_A$
up to a factor of 2 more than the contributions from
\cite{burrows:98}. We refer the reader to Figure 2 of
\cite{horowitz:17} for a graphical display of $S_A$ and $S^f_A$ for different
$\rho$, $T$, and $Y_e$ values and for various neutrino interaction
assumptions.

The goal of these proceedings is to explore the effect of these Virial
contributions, as well as the many-body effects in general, on the
core-collapse supernova dynamics. To this end, we perform a large
number of simulations, broken into two parts. First, we simulate
core-collapse in four progenitor models from \cite{woosley:07}:
s12WH07, s15WH07, s20WH07, and s25WH07. For each progenitor we run a
simulation with and without the corrections from \cite{horowitz:17}.
Second, for model s20WH07 we simulate an additional 15 simulations,
split into three sets: without corrections, with only high density
corrections (i.e. $B_0=0$), and with full corrections. In an attempt
to remove the effect of stochastic motions on our results and
interpretations, for each of the three sets we do five simulations,
each one starting with different random perturbations (at the 0.1\%
level) to the density field. We apply these perturbations everywhere
at the start of the 2D simulation. Each simulation is started at
15\,ms after bounce. We take the evolution up to 15\,ms from a 1D {\tt
  GR1D} simulation \citep{oconnor:15a}.

\section{Results}

\subsection{Progenitor Dependence}

\begin{figure}[tbh]
\centering
 \includegraphics[width=5in]{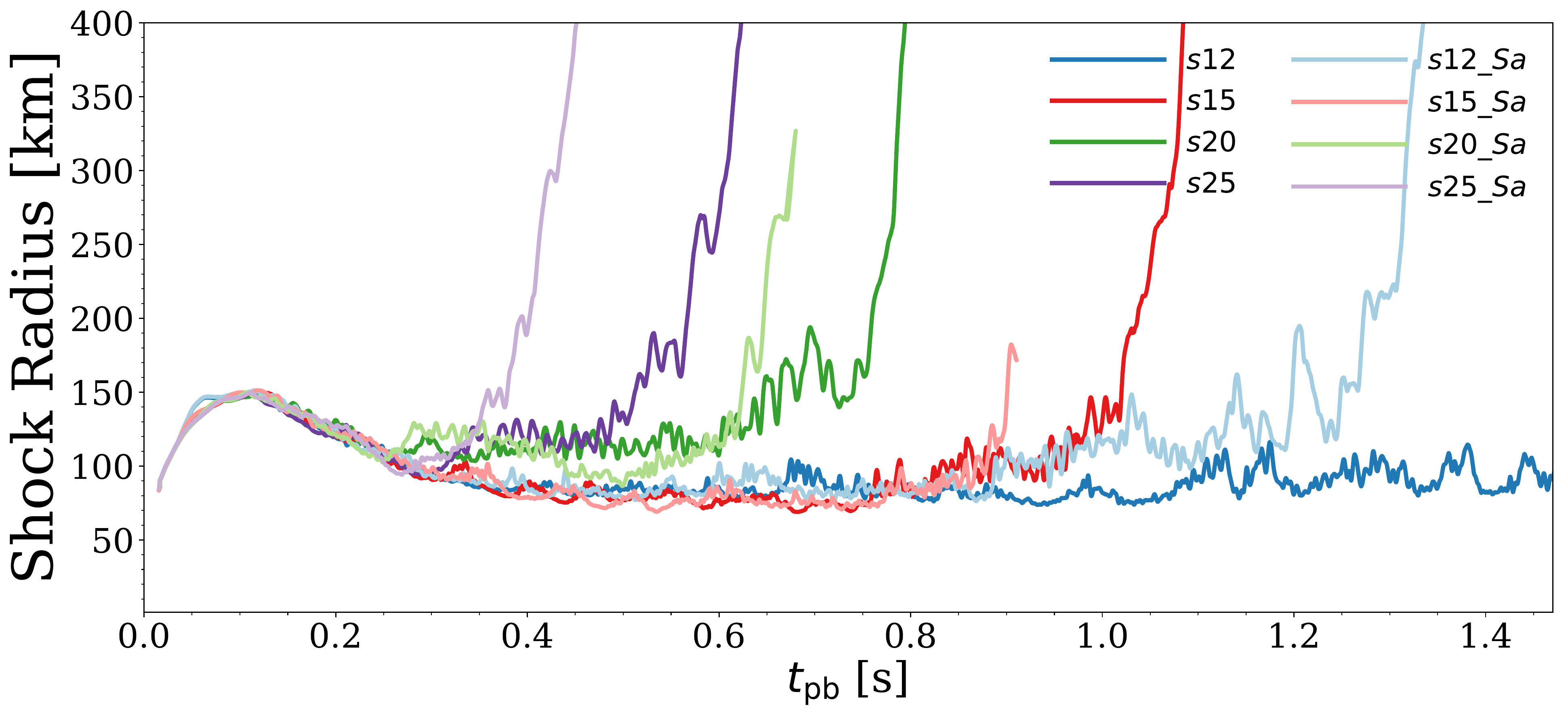} 
 \caption{Shock radius evolution for four different progenitors, with
   and without the explored  corrections
   from \cite{horowitz:17}. The corrections lead to earlier
   explosions.}
   \label{fig:progenitors}
\end{figure}

We simulate core-collapse in four progenitors, with and without the
neutral-current scattering cross section corrections discussed above.
We show the evolution of the shock radius in
Figure~\ref{fig:progenitors}.  Dark colours represent the simulations
without the corrections, while the lighter colours are the
corresponding simulations with corrections.  In all cases the shock
radius propagates out to $\sim 150$\,km, stalls, and recedes. For
models s15WH07, s20WH07, and s25WH07, the reduced neutral-current
scattering cross sections lead to earlier explosions. Explosions times
are reduced by $\sim$100\,ms -150\,ms.  For model s12WH07, the
corrections lead to a late-time explosion where there was not one
observed in the model without corrections.  The impact of these
corrections is more modest than determined by \cite{burrows:16}, where
the inclusion of these corrections dramatically decreased the
explosion times in several models.

\subsection{s20WH07}

\begin{figure}[tbh]
\centering
 \includegraphics[width=5in]{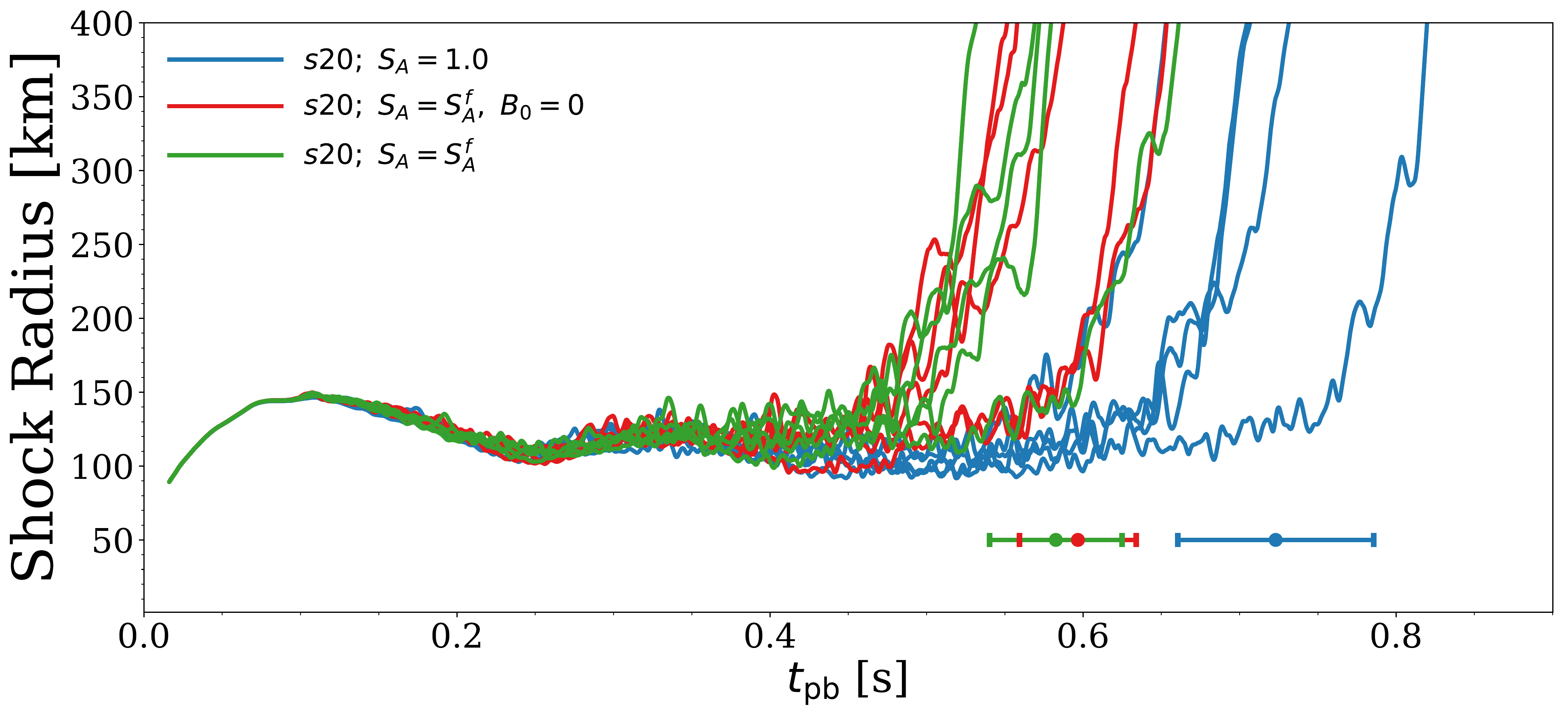} 
 \caption{Shock radius evolution for 15 simulations of s20WH07 using
   three different neutrino-nucleon neutral current scattering cross
   sections based on Eq.~\ref{eq:sigma}.  Five use $S_A=1$ (blue),
   five take $S_A = S_A^f; B_0=0$ (red), the
   remaining five use $S_A = S_A^f$ (green). We include
   the average explosion time and the variance within the five
   simulations.}
   \label{fig:15shocks}
\end{figure}

Two dimensional simulations of core-collapse supernovae are sensitive
to stochastic variations in the matter motions during the accretion
phase. This can lead to differences in explosion times which could
potentially cloud interpretations. To overcome this, and to probe the
effect of the corrections discussed in \S~\ref{sec:methods} more
deeply, we perform many simulations with the s20WH07 model. As
mentioned above, we run 15 simulations in total and explore three
unique setups. Each simulation is started with a different set of
random perturbations on the density field
($\rho \to \rho \times (1+r)$; where $r$ is a different random number
for each zone between -0.001 and 0.001). The first set of five
simulations ignores any corrections to the neutral-current scattering
cross section from the axial response, i.e. $S_A=1$ in
Eq.~\ref{eq:sigma}. The second set of simulations uses $S_A$ as shown
in Eq.~\ref{eq:S_A} but sets $B_0=0$ and the third set used
Eq.~\ref{eq:S_A} directly. We show the evolution of the shock radius
in these 15 simulations in Figure~\ref{fig:15shocks}. We colour-code
each set, but do not distinguish between the five runs of each set for
clarity. We see the same results as the previous section. The reduced
value of $S_A$ due to many-body effects leads to earlier explosions.
We define the explosion time to be when the shock crosses
400\,km. With this definition, the average explosion times, and the
variances are determined to be:
$$t_\mathrm{exp}^{S_A=1} = (723\pm63)\,\mathrm{ms};\ \ \ \ \  t_\mathrm{exp}^{S_A^f, B_0=0} = (597\pm37)\,\mathrm{ms};\ \ \ \ \  t_\mathrm{exp}^{S^f_A} = (583\pm42)\,\mathrm{ms}\,.$$
The key observation we make here is that the Virial EOS contributions
to the axial response, which lower $S_A$ at lower densities,
do not have a significant effect on the explosion dynamics.  Statistically the
explosion times are indistinguishable from those that only include the
reduction due to the effective RPA corrections.  We explore this in
more detail by examining the neutrino quantities.

\begin{figure}[tbh]
\centering
 \includegraphics[width=0.45 \textwidth]{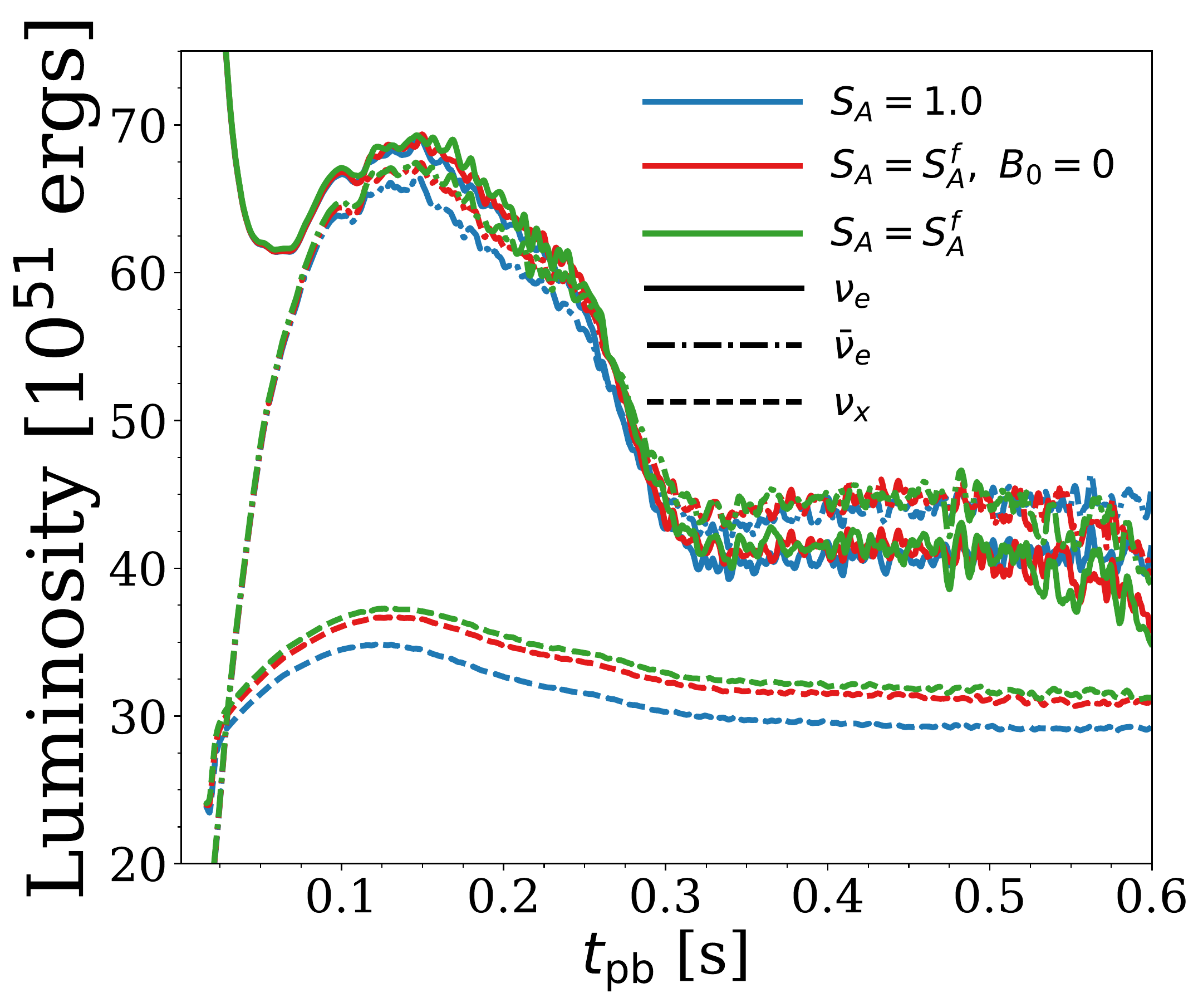} 
 \includegraphics[width=0.45 \textwidth]{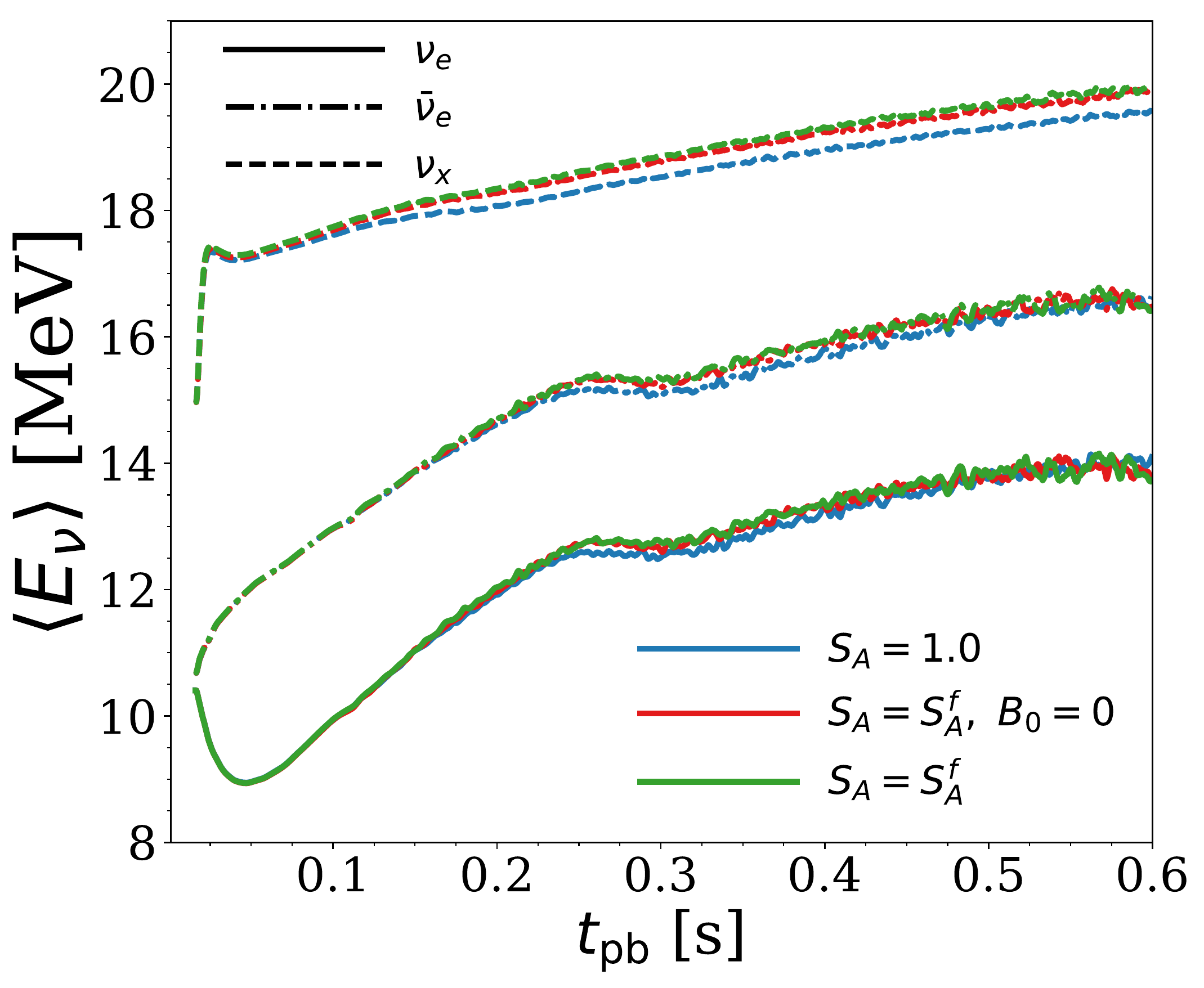} \\
 \includegraphics[width=0.45 \textwidth]{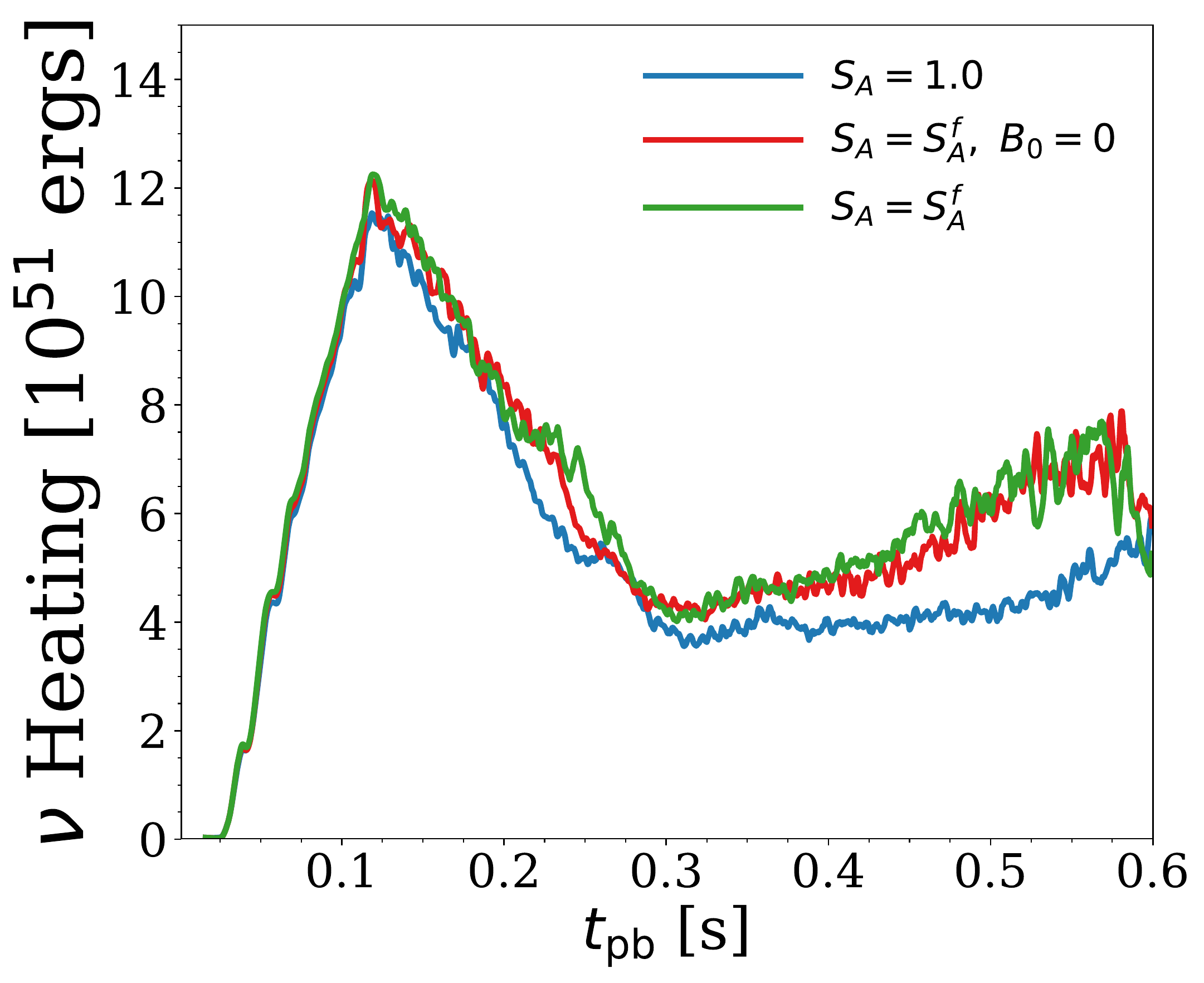} 
 \includegraphics[width=0.45 \textwidth]{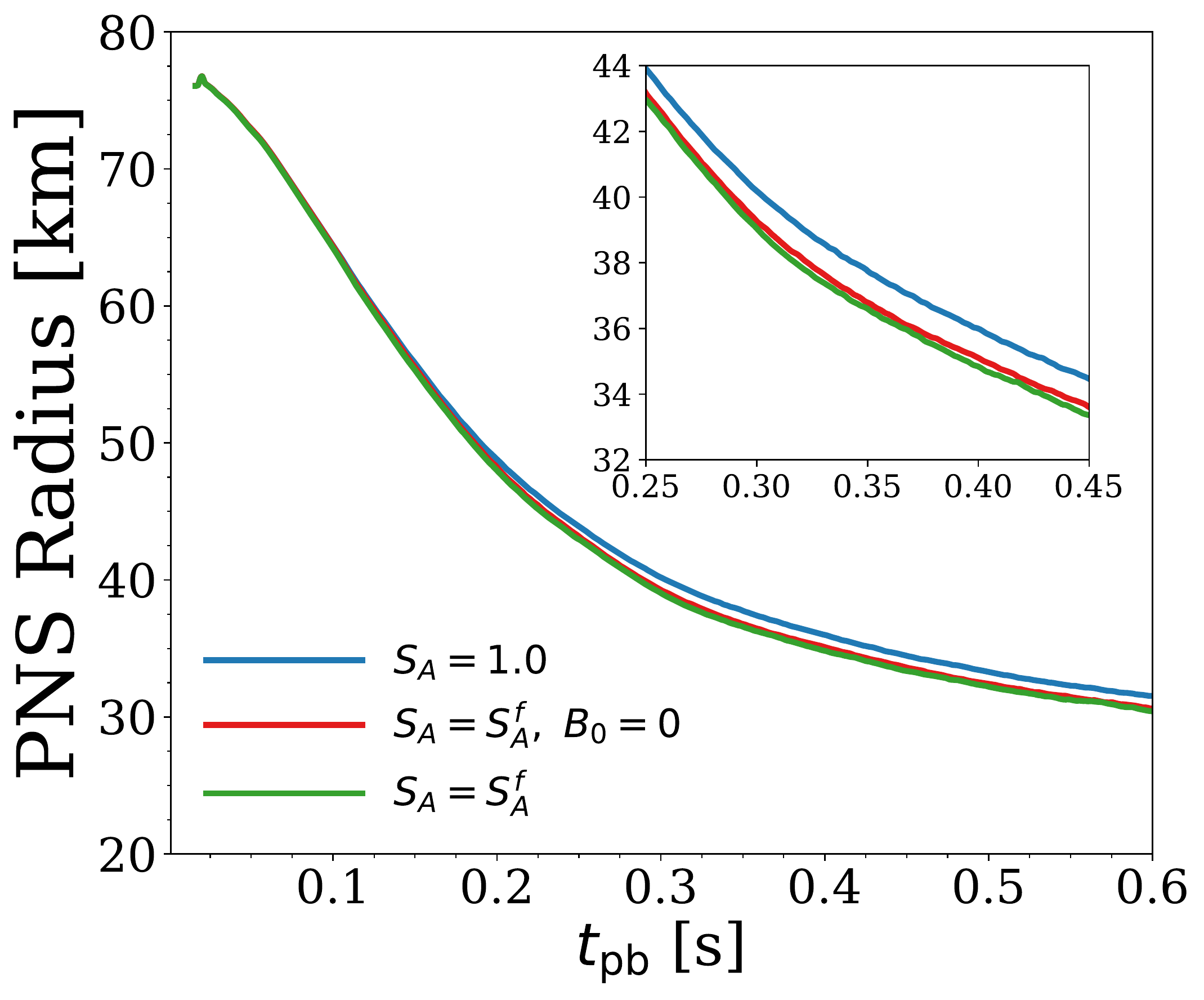} 
 \caption{Neutrino related quantities for model s20WH07.  Here we show the neutrino
   luminosity (top left) and average energy (top right) for each
   neutrino species.  We also show the net neutrino heating (bottom
   left) and the protoneutron star radius (bottom right; defined as
   the radius of $\rho=10^{11}$\,g\,cm$^{-3}$).  For each of the three
 configurations studied, we average the quantities from the five
 simulations in order to reduce scatter and better ascertain the
 overall effect.}
   \label{fig:neutrinoquants}
\end{figure}

In Figure~\ref{fig:neutrinoquants}, we show graphs of various neutrino
quantities over the first 600\,ms of evolution for the three different
sets of simulations. For each set we average the quantities for the
five different simulations. This significantly reduces the scatter and
allows us to see the effect of each neutrino opacity change in detail.
In the upper left panel we show the luminosity for each of the three
neutrino species and for each set of simulations. The largest effect
we see is the increase in the $\nu_x$ luminosity, by $\sim$10\%, when
the many-body corrections to the axial response function are
considered. The lower total scattering rates that result from the
lower $S_A$ allow the $\nu_x$ neutrinos to escape more easily and
results in more cooling. Consequently, this leads to higher $\nu_x$
average energies (upper right panel) and faster contracting
protoneutron star radii (bottom right panel). The faster contraction
has the side effect of increasing the electron type neutrino and
anti-neutrino luminosities and average energies which leads to
increased neutrino heating (bottom left panel). This is the same
effect observed in \cite{melson:15b} when they reduced the
neutral-current scattering by considering relatively
large strange-quark contributions.

Here we notice that the Virial contributions do indeed lead to more
cooling, higher energies, and faster contraction compared to the $B_0=0$
simulation, as expected. However, the additional effect is relatively
small compared to the impact the Virial contributions have on $S_A$
(upwards of a factor of 2 more important than the RPA corrections at
low densities). We can resolve this apparent conflict by considering
the properties of the region where the $\nu_x$'s are emitted from and
scatter through and comparing those to the region where the Virial
contributions dominate. Unlike $\nu_e$ and $\bar{\nu}_e$, $\nu_x$ do
not undergo charged-current interactions. The dominant emission
process is via pair production. Due to the temperature dependence, the
luminosity build up occurs deeper (and at higher densities,
$\rho \sim 10^{12}-10^{13}$\,g\,cm$^{-3}$, and temperatures
$T\sim10\,$MeV) in the protoneutron star. At these densities and
temperatures the corrections to $S_A$ in Eq.~\ref{eq:S_A}, which are
of order $\sim$20\%, are either dominated by the effective RPA (at the
highest densities) or of similar value between the Virial and
effective RPA. These two reasons explain why the Virial EOS
contributions to the axial response do not significantly affect the
core-collapse supernova dynamics.

\section{Conclusions}

In this proceedings for IAUS 331, we present a detailed look at the
effect of corrections to the neutral-current neutrino-nucleon
scattering cross sections on the core-collapse supernova explosion
dynamics. In particular, we look at recent model-independent
corrections to the axial response ($S_A$) of low-density nuclear
matter described by the Virial EOS. We argue here that corrections to
$S_A$ at the level of $\sim$20\% can
influence the dynamics, however, at the relevant densities
($10^{12}-10^{13}$\,g\,cm$^{-3}$) and temperatures ($\sim$10\,MeV) our
model-independent calculation begins to break down and we must resort
to model-dependent, RPA calculations to estimate $S_A$. Due to the
important role these corrections may play in the evolution of the
central engine of core-collapse supernovae, we advocate for a better
understanding of the neutrino response at these densities.

\emph{Acknowledgements}: We thank O.L. Caballero and A. Schwenk for
collaborative work on the Virial EOS. Support was provided by NASA through Hubble Fellowship grant
\#51344.001-A awarded by the STScI, which is operated by the AURA, for
NASA, under contract NAS 5-26555. Computations were performed on the
Zwicky cluster at Caltech, supported by the Sherman Fairchild
Foundation and by NSF award PHY-0960291, and on XSEDE which is
supported by National Science Foundation grant number ACI-1548562.


\begin{thebibliography}{}

\bibitem[{Bruenn}, 1985]{Bruenn:85}
{Bruenn}, S.~W. (1985).
\newblock {\em \apjs}, 58:771--841.

\bibitem[{Burrows} and {Sawyer}, 1998]{burrows:98}
{Burrows}, A. and {Sawyer}, R.~F. (1998).
\newblock {\em \prc}, 58:554.

\bibitem[{Burrows} et~al., 2016]{burrows:16}
{Burrows}, A., {Vartanyan}, D., {Dolence} \emph{et al.}(2016).
\newblock {\em ArXiv e-prints: 1611.05859}.

\bibitem[{Couch} and {O'Connor}, 2014]{couch:14a}
{Couch}, S.~M. and {O'Connor}, E.~P. (2014).
\newblock {\em \apj}, 785:123.

\bibitem[{Fryxell} et~al., 2000]{fryxell:00}
{Fryxell}, B., {Olson}, K., {Ricker}, P., \emph{et al.}.
  (2000).
\newblock {\em \apjs}, 131:273.

\bibitem[{Horowitz}, 2002]{horowitz:02}
{Horowitz}, C.~J. (2002).
\newblock {\em \prd}, 65(4):043001.

\bibitem[{Horowitz} et~al., 2017]{horowitz:17}
{Horowitz}, C.J., {Caballero}, O.L., {Lin}, Z., \emph{et al.}. (2017).
\newblock {\em \prc}, 95(2):025801.

\bibitem[{Melson} et~al., 2015]{melson:15b}
{Melson}, T., {Janka}, H.-T., {Bollig}, R., {Hanke}, F., {Marek}, A., and
  {M{\"u}ller}, B. (2015).
\newblock {\em \apjl}, 808:L42.

\bibitem[{O'Connor}, 2015]{oconnor:15a}
{O'Connor}, E. (2015).
\newblock {\em \apjs}, 219:24.

\bibitem[{O'Connor} and {Couch}, 2015]{oconnor:15b}
{O'Connor}, E. and {Couch}, S.~M. (2015).
\newblock {\em submitted to \apj; arXiv:1511.07443}.

\bibitem[{Lattimer} and {Swesty}, 1991]{lseos:91}
{Lattimer}, J.~M. and {Swesty}, F.~D. (1991).
\newblock {\em Nuc. Phys.}, A535:331.

\bibitem[{Woosley} and {Heger}, 2007]{woosley:07}
{Woosley}, S.~E. and {Heger}, A. (2007).
\newblock {\em \physrep}, 442:269.

\end{thebibliography}
\end{document}